# Spin ice behavior in $Dy_2Sn_{2-x}Sb_xO_{7+x/2}$ and $Dy_2NbScO_7$

X. Ke[1*], B. G. Ueland[1*†], D.V. West[2], M. L. Dahlberg[1], R. J. Cava[2], and P. Schiffer[1]

[1]*Department of Physics and Materials Research Institute, Pennsylvania State University, University Park PA 16802*

[2]*Department of Chemistry and Princeton Materials Institute, Princeton University, Princeton, NJ 08540*

We report the magnetic and thermal properties of $Dy_2Sn_{2-x}Sb_xO_{7+x/2}$, $x$ = 0, 0.25, and 0.5, and $Dy_2NbScO_7$. We find evidence for Ising-like single ion ground states in the $Dy_2Sn_{2-x}Sb_xO_{7+x/2}$ materials. These materials possess nearly the same zero point entropy as the canonical spin ices $Ho_2Ti_2O_7$ and $Dy_2Ti_2O_7$, strongly suggesting that they have spin ice states at low temperatures. We also observe a somewhat reduced zero point entropy in $Dy_2NbScO_7$, which is possibly associated with the higher level of cation disorder. The ice-like states in these materials with cation disorder on the B-sites of the pyrochlore lattice provide new evidence for the robust nature of spin ice behavior in the presence of disorder.

In geometrically frustrated magnetic materials, the geometry of a regular magnetic lattice results in close competition between local spin-spin interactions. Such frustration can lead to a large manifold of nearly degenerate spin states at low temperatures, creating a wide range of novel low temperature behavior including spin-liquid-like states, exotic long-range-ordered states, and frozen states, which show spin-glass-like behavior in the presence of minimal disorder [1,2,3,4]. One class of geometrically frustrated magnets that displays such frozen disordered magnetic states has been named "spin ice" [3,5,6], due to the close analogy between its local magnetic correlations and the local $H^+$ ordering in water ice.

In water ice, two possible O-H bond lengths (long and short) combined with the tetrahedral coordination of four $H^+$ ions around each $O^{2-}$ ion lead to macroscopic configurational degeneracy [7] associated with two short and two long bonds around each $O^{2-}$ ion. This degeneracy manifests itself in experiments as a "zero point entropy," meaning that the low temperature macroscopic state of the system has entropy associated with the disordered configuration of the $H^+$ ions [8]. Pauling showed that the entropy should fall short of that expected for a two state system ($S$ = Rln 2), resulting in a total entropy of $S$ = R[ln 2 – (1/2) ln(3/2)], where R is the gas constant [9]. Spin ice behavior is associated with the rare earth pyrochlore materials ($A_2B_2O_7$), in which the magnetic rare earth ions (A) are situated on the vertices of a lattice of corner sharing tetrahedra. Due to the local crystalline fields, the magnetic ground states of these ions are Ising-like, pointing along an axis joining the center of two neighboring tetrahedra [5, 6, 10]. The balance between the dipolar and exchange interactions between spins result in a ground state for each magnetic tetrahedron with two spins pointing in and two pointing out. This local spin configuration is ice-like, i.e., it is analogous to the two different bond states for the O-H bonds in water ice, and it leads to the same statistical mechanics for the collective spin states and hydrogen states in water ice, including the existence of zero point entropy [6, 11]. Although recent numerical calculations [12] suggest the existence of a long range ordered state for spin ice systems, current experimental data on the spin ice materials show that the spins freeze into a disordered low temperature state, and no long range ordering has yet been observed experimentally in zero magnetic field down to the lowest temperatures measured [5, 13, 14].

As stated above, the spin ice state has been confirmed in the two canonical spin ice materials $Ho_2Ti_2O_7$ [5, 13] and $Dy_2Ti_2O_7$ [14, 15]. Additionally, recent neutron scattering experiments indicate that $Ho_2Sn_2O_7$ belongs to the spin ice family as well [16]. Magnetic susceptibility measurements on $Dy_2Sn_2O_7$ have also indicated that this material is a candidate spin ice material, due to the similarities of the temperature and frequency dependence of its ac susceptibilities to those of the titanate spin ices [17]. To date, however, no measurements of the local ordering or zero point entropy have been utilized to probe its low temperature state. In this report, we present magnetization, ac susceptibility, and specific heat measurements on $Dy_2Sn_{2-x}Sb_xO_{7+x/2}$, $x$ = 0, 0.25, and 0.5. Our magnetization data suggest that the magnetic $Dy^{3+}$ cations in these materials have doublet ground states with strong single ion anisotropy, causing them to behave like Ising spins. The magnetic entropy determined from heat capacity measurements also confirms that all three have zero-point entropy consistent with a disordered spin ice state at low temperatures. These results expand the class of confirmed spin ice materials to include compounds where the pyrochlore B site (e.g., the site on which the $Ti^{4+}$, $Sn^{4+}$, or $Sb^{3+}$ cation is located) has a mixed valence. We also probe the effect of B site cation disorder on the spin relaxation behavior and low temperature state by characterizing another compound, $Dy_2NbScO_7$, which also appears to have an ice-like low temperature state, although with somewhat less zero point entropy.

Polycrystalline samples of $Dy_2Sn_{2-x}Sb_xO_{7+x/2}$, $x$ = 0, 0.25, and 0.5 and $Dy_2NbScO_7$, were synthesized using standard solid state synthesis techniques and were determined to be single phase pyrochlores by X-ray diffraction. We measured the magnetization $M$ of each sample using a Quantum Design superconducting quantum interference device (SQUID) magnetometer. The specific heat $c_p$ and ac magnetic susceptibility $\chi = dM/dH$ were determined using a Quantum Design Physical Property Measurement System (PPMS). Specific heat data were taken down to $T = 0.4$ K with the $^3$He option of the PPMS using a standard semi-adiabatic heat pulse technique, and the samples measured were mixed with Ag powder and pressed into pellets to facilitate thermal equilibration (all data shown have the addendum from the Ag and sample holder subtracted). The real and imaginary parts of the ac susceptibility, $\chi'$ and $\chi''$, were determined using an excitation field of $H_{ac}$ = 1 or 10 Oe at frequencies spanning $f$ = 10 – 10000 Hz. The data were independent of the magnitude of the excitation field used.

In Fig. 1, we show the field dependence of magnetization at $T$ = 1.8 K for $Dy_2NbScO_7$ and the three $Dy_2Sn_{2-x}Sb_xO_{7+x/2}$ materials. All display similar behavior, having a saturated magnetization $M_{sat} \approx 5$ $\mu_B$/Dy, which is in good agreement with previously published data on the titanate spin ices [18,19]. These data suggest that the $Dy^{3+}$ ions in these materials possess an Ising type single ion ground state for all three compositions. The slope of $M(H)$ at large fields increases slightly with increasing $x$, and is noticeably larger for $Dy_2NbScO_7$. This suggests that B-site disorder may affect the crystal electric field and thus the Ising-like nature of the spins. The inset to Fig. 1 shows the inverse static susceptibility $\chi_{dc}^{-1} = H/M$ of all three samples determined from $M(T)$ data taken in a field of $H$ = 0.02 T. Data were fit to the Curie-Weiss law between $T$ = 10 – 20 K to determine the effective magnetic moment $p$ and the Weiss temperature $\theta_W$. We found that $p$ remains near the expected value for a free $Dy^{3+}$ ion of $p$ = 10.6 $\mu_B$ for all samples [20], and that $\theta_W(x)$ monotonically decreases from $\theta_W$ = 0.19 K at $x$ = 0 to $\theta_W$ = -0.32 K at $x$ = 0.5. This small variation suggests that the effective magnetic interactions in the material become increasingly antiferromagnetic as $Sb^{5+}$ replaces $Sn^{4+}$.

Figure 2 shows the ac susceptibility versus temperature for all three compositions of $Dy_2Sn_{2-x}Sb_xO_{7+x/2}$ and $Dy_2NbScO_7$, in which the main panels and insets show data for $\chi'$ and $\chi''$, respectively. All three samples show Curie type behavior for $T$ > 30 K, in which $\chi'$ monotonically increases with decreasing temperature while $\chi''(T) \sim 0$, for all of our measurement frequencies. At $T \approx 25$ K and $T \approx 2$K, frequency dependent maxima appear in both $\chi'$ and $\chi''$, indicating that spin freezing occurs at these temperatures on the time scale of the measurements. The high temperature maxima in the susceptibility is the most pronounced in the data for $Dy_2Sn_2O_7$ (Fig. 2a), and the data are consistent with the previously published results [17]. This behavior is likely due to the same type of single ion freezing seen in similar data on $Dy_2Ti_2O_7$ [19,21] and seen in higher field data for $Ho_2Ti_2O_7$ [22]. As shown in Fig. 2b and 2c, the high temperature spin freezing is less pronounced for the $x$ = 0.25 and 0.5 samples, meaning that the higher temperature maxima in $\chi(T)$ appear at higher frequencies with increasing $x$, indicating that the characteristic spin relaxation time becomes faster in samples for which more $Sb^{5+}$ replaces $Sn^{4+}$. The dependence of the spin relaxation time on composition also manifests itself in the low temperature peak in $\chi$, which corresponds to the cooperative freezing associated with the spin ice state in $Dy_2Ti_2O_7$ and $Ho_2Ti_2O_7$. That both the high temperature and low temperature spin freezing are only seen at higher frequencies for the $x = 0.25$ and $0.5$ samples, and that both are absent in $Dy_2NbScO_7$ within our measurement frequency window, as shown in Fig. 2(d), indicate that the spin relaxation time is shorter in these materials, as is demonstrated more explicitly below.

To further investigate the change in the spin relaxation time with $x$, we measured $\chi(f)$ at constant temperature. Figure 3a shows an example of data collected on $Dy_2Sn_2O_7$, in which we plot $\chi''(f)$ at $H = 0$ for different temperatures. Following the same procedure as in previous studies on pure and non-magnetically diluted samples of $Dy_2Ti_2O_7$ [19,21], we determined the characteristic spin relaxation time $\tau$ at each temperature from the inverse of the frequency at which a maximum in $\chi''(f)$ occurred. These results are shown in Fig. 3b, where we plot $\tau(T)$ for all three $Dy_2Sn_{2-x}Sb_xO_{7+x/2}$ samples as well as previously published data on $Dy_2Ti_2O_7$ [19] for comparison. For $Dy_2Ti_2O_7$, $\tau$ is strongly temperature dependent at the lowest and highest temperatures in our measurement window but is practically temperature independent between $T \approx 5$-$13$ K. In our previous work, we attributed this behavior to the dominant spin relaxation mechanism changing from thermal activation to quantum tunneling, and then back to thermal activation with cooling [19]. Interestingly, in the three $Dy_2Sn_{2-x}Sb_xO_{7+x/2}$ samples studied here, we observe similar qualitative behavior, suggesting that the spin relaxation mechanisms are quite similar. Arrhenius fits of the form $f = f_0 e^{-E_A/K_BT}$, used to describe simple thermally activated relaxation over an energy barrier $E_A$, were performed to the temperature dependence of the characteristic frequency, $f = 1/\tau$, on data for $Dy_2Ti_2O_7$ and $Dy_2Sn_2O_7$ in the high temperature region of our measurement window. For $Dy_2Ti_2O_7$, fits between $T = 15 - 17$ K yield an activation temperature of $E_A/k_B \approx 190$ K, which corresponds well to the known first excited crystal field state of the material [19,23]. Similarly, we find $E_A/k_B \approx 220$ K for $Dy_2Sn_2O_7$ between $T = 19 - 23$ K. This is perhaps not surprising, since one may expect that the similar symmetry and distances within the unit cell of the crystal lattice should produce crystal field splitting of the $Dy^{3+}$ multiplet similar to that seen in $Dy_2Ti_2O_7$. In addition, the fact that the activation energies are so similar means that the size and chemical differences between Ti and Sn have little effect on the crystal field splitting. This contrasts with data on the non-magnetic diluted materials $Dy_{2-x}Y_xTi_2O_7$, in which $E_A$ dramatically increases with increasing dilution [19]. However, it is noteworthy that the attempt frequency $f_0$, which can be thought of as how often the spins try to cross the energy barrier $E_A$, is about $f_0 \approx 100$ MHz for $Dy_2Sn_2O_7$, which is almost a factor of 5 smaller than the value found for $Dy_2Ti_2O_7$ [19].

While the spin relaxation processes in $Dy_2Sn_2O_7$ appear slower than those in $Dy_2Ti_2O_7$, we see in Fig. 3b that the substitution of $Sb^{5+}$ for $Sn^{4+}$ dramatically decreases the spin relaxation time at a given temperature. This is similar to previous observations on $Dy_{2-x}Y_xTi_2O_7$ in the limit of small $x$ [19], and also to observations on $Ho_{2-x}Y_xTi_2O_7$ for large $x$, though under high magnetic fields [24]. These observations have been attributed to an increasing value of $E_A$ with dilution. However, we find that the values of $E_A$ determined from our Arrhenius fits to the data for $Dy_2Sn_{2-x}Sb_xO_{7+x/2}$ decrease considerably in the high temperature region with the substitution of $Sb^{5+}$ for $Sn^{4+}$, with $E_A/k_B \approx 42$ K for $x = 0.25$, coincident with a decrease in $f_0$ to be on the order of 50 kHz.

We find that $E_A$ appears to be even smaller for $Dy_2Sn_{1.5}Sb_{0.5}O_{7.25}$, although we could not extract the exact value of $E_A$ and the attempt frequency from the Arrhenius fit due to the limits of our measurement window. Since the difference of the lattice parameters of these three samples is negligible, we believe that there must be other mechanisms affecting the crystalline electric field, which subsequently determines the energy barriers, and hence the fitted value of $E_A$. One candidate is the resultant disorder from the mixed occupancy of the B-site elements, but more detailed studies of the spin relaxation mechanism as a function of Sb doping would be required to demonstrate this to be the case. To gather additional insight into the effect of B site disorder on the spin relaxation process, we also measured the frequency dependence of the ac susceptibility for $Dy_2NbScO_7$. As expected from Fig. 2(d), we do not observe any maxima in the $\chi''(f)$ curves for temperatures and frequencies within our measurement window. This indicates that the spin relaxation process in this material is much faster than that observed in $Dy_2SnO_7$ and $Dy_2Ti_2O_7$.

Due to the similarity in the magnetizations and ac susceptibilities of $Dy_2Sn_{2-x}Sb_xO_{7+x/2}$ and $Dy_2Ti_2O_7$, it is reasonable to expect that the $Dy_2Sn_{2-x}Sb_xO_{7+x/2}$ materials freeze into a spin ice state. To confirm this hypothesis, we have obtained the magnetic entropy $S$ from heat capacity measurements by integrating $c_{mag}/T$, where $c_{mag}$ is the magnetic contribution to the total heat capacity, from the lowest temperatures measured. The difference between the integrated entropy and the expected value for an Ising system ($S = R\ln 2$) is the zero point entropy [13, 14], and this difference constitutes the primary thermodynamic evidence for the existence of a spin ice state. Figure 4a shows our specific heat data for $Dy_2Sn_2O_7$ and $Dy_2Sn_{1.5}Sb_{0.5}O_{7.25}$ (i.e., $x = 0.5$) at $H = 0$ and 1 T. We subtracted out the phonon contributions to the specific heat by fitting the data between $T = 10 - 20$ K to the Debye model, and plot the resulting magnetic specific heat in the inset of Fig. 4a as $c_{mag}(T)/T$, where we have included data on $Dy_2Ti_2O_7$ (the black curves in the figure) for comparison. Figure 4b shows the integrated magnetic entropy at $H = 0$ and 1 T for both $Dy_2Sn_2O_7$ and $Dy_2Sn_{1.5}Sb_{0.5}O_{7.25}$. We find that $S(T)$ increases with increasing temperature, leveling off at the spin ice value of $S = 0.71R\ln2$, for both materials, at $H = 0$. This is identical to what has been observed in the well known spin ices, $Dy_2Ti_2O_7$ [14] and $Ho_2Ti_2O_7$ [13], confirming that $Dy_2Sn_2O_7$ has a spin ice state at low temperatures. As previously seen in the titanate spin ices [13,14], application of a magnetic field results in the total integrated entropy increasing towards the expected value for an Ising system of $S = R\ln2$, thus reducing the zero point entropy. We note that for the $x = 0.5$ material, the inset to Fig. 4a shows a broader peak in $c_{mag}(T)$, indicating that more entropy may be recovered at lower temperatures in $Dy_2Sn_{1.5}Sb_{0.5}O_{7.25}$, and indeed crudely extrapolating this curve to $T = 0$ by a straight line yields a 12% increase in $S$ at $H = 0$ and a 4% in $S$ at $H = 1$ T. However, these differences amount to only a fraction of the total zero point entropy, and are likely the result of the disorder associated with the mixture of $Sn^{4+}$ and $Sb^{5+}$ ions and the resulting oxygen placement in the lattice. The small size of this deviation from the behavior of $Dy_2Ti_2O_7$ and $Dy_2Sn_2O_7$ suggests that the local spin correlations in the materials with Sb substitution are still largely spin-ice-like, despite the presence of the disorder and the dramatic decrease of $E_A$ previously discussed. The data in Fig. 4(b) show that the integrated entropy of $Dy_2NbScO_7$ is larger than the expected value for a spin ice system. This indicates that the disorder on B site could break the strict 'ice rule' for the arrangements of the $Dy^{3+}$ spins on the A sites of the pyrochlore structure, leading to a smaller zero point entropy. Recently, a decrease of zero point entropy of another pyrochlore material, $Ho_2Ru_2O_7$, has also been reported, which is attributed to the internal field of the order of 1 T associated with the presence of magnetic $Ru^{4+}$ ions [25].

In summary, our data confirm that $Dy_2Sn_2O_7$ is a spin ice and strongly suggest that the $x$ = 0.25 and 0.5 materials are likely spin ice materials with local Ising moments and ice-like zero point entropy, while even $Dy_2NbScO_7$ has considerable zero point entropy suggesting an ice-like character to the low temperature state. These findings increase the number of known spin ice systems, expanding the category to include materials in which the B-site has mixed valence. That the spin ice state (as indicated by the zero point entropy) is relatively unchanged by introducing disorder is perhaps not surprising, given recent findings on the "stuffed" spin ice system [26], and the results open up the possibility of theoretical studies of the reaction of the spin ice state to a range of different types of structural disorder.

We acknowledge the financial support from NSF grant DMR-0353610.

**FIGURE CAPTIONS**

Figure 1. The magnetic field dependence of the magnetization $M$ at $T = 1.8$ K for $Dy_2Sn_{2-x}Sb_xO_{7+x/2}$ with $x = 0$, 0.25, 0.5, and $Dy_2ScNbO_7$ (dark yellow curve). The inset shows inverse dc magnetic susceptibility $\chi^{-1}$ versus temperature with straight lines representing Curie-Weiss fits between $T = 10$ K and 20 K.

Figure 2. The measured ac magnetic susceptibility versus temperature for $Dy_2Sn_{2-x}Sb_xO_{7+x/2}$ for $x = 0$ (a), 0.25 (b), 0.5 (c), and $Dy_2ScNbO_7$ (d), at $f = 10$, 100, 1000, and 10000 Hz. The main panels show the real part of ac susceptibility $\chi'$ and the insets show the imaginary part $\chi''$.

Figure 3. (a) Frequency dependence of $\chi''$ for $Dy_2Sn_2O_7$ at different temperatures in zero applied field. (b) Temperature dependence of the characteristic spin relaxation time $\tau$ for $Dy_2Ti_2O_7$, and $Dy_2Sn_{2-x}Sb_xO_{7+x/2}$ with $x = 0$, 0.25, 0.5. Data on $Dy_2Ti_2O_7$ has been previously published in Ref.[19].

Figure 4. (a) The specific heat versus temperature for $Dy_2Sn_2O_7$ and $Dy_2Sn_{1.5}Sb_{0.5}O_{7.25}$ at $H = 0$ and 1 T before subtracting out the lattice contributions. Black curves represent data for $Dy_2Ti_2O_7$. The magnetic heat capacity divided by $T$ is shown in the inset. (b) The integrated entropy as a function of temperature. The integrated entropy of $Dy_2ScNbO_7$ at $H = 0$ T is represented by the dark yellow curve.

Figure 1.
X. Ke, et al.

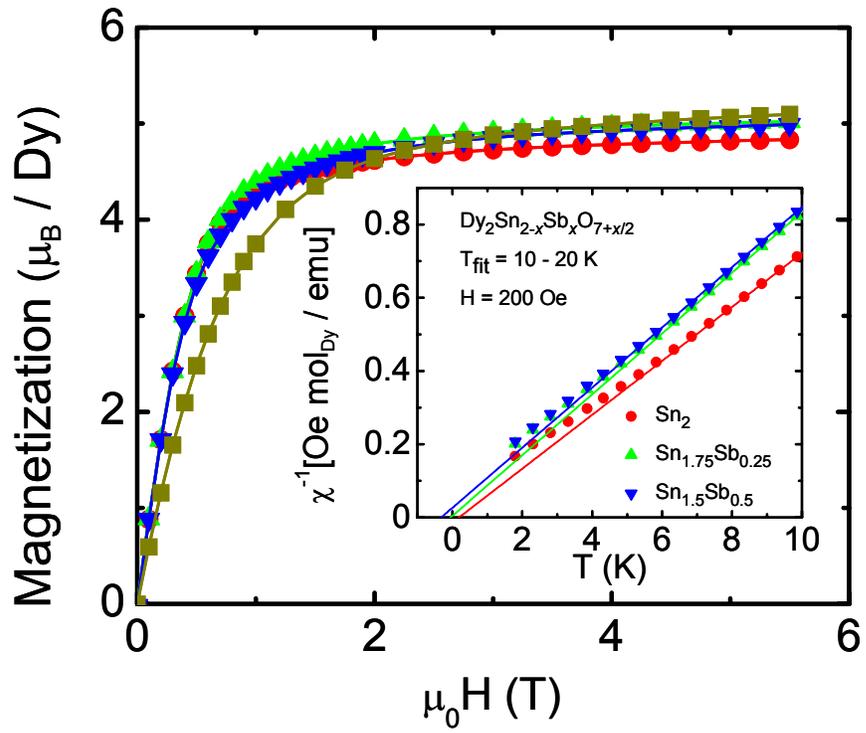

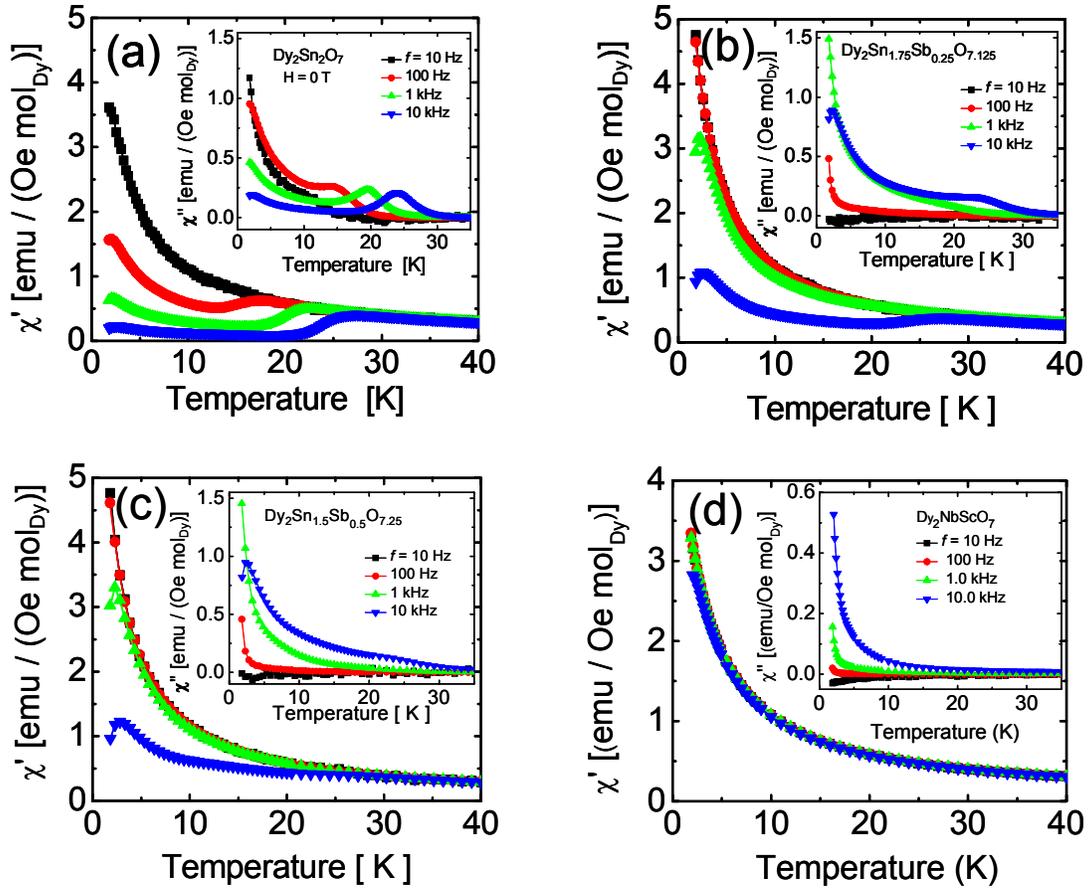

Figure 3.
X. Ke, et al.

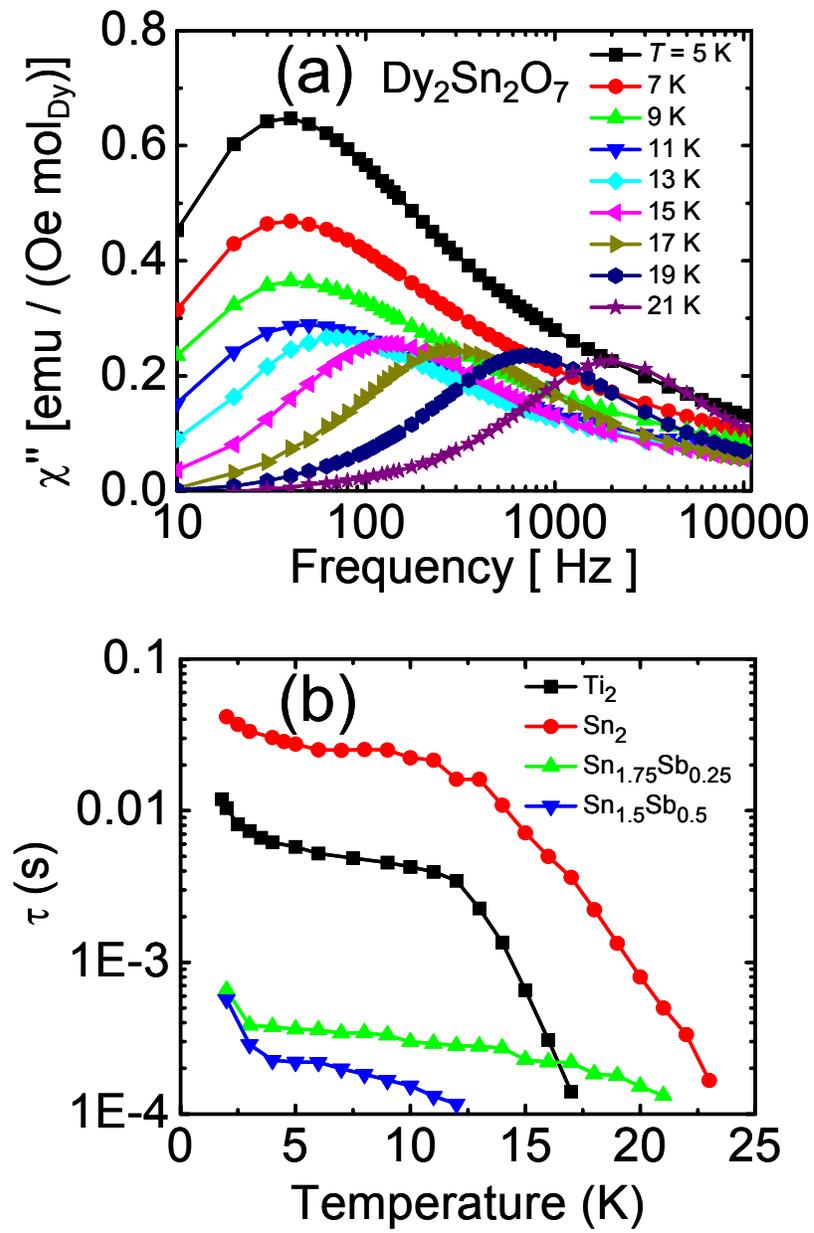

Figure 4.
X. Ke et al,

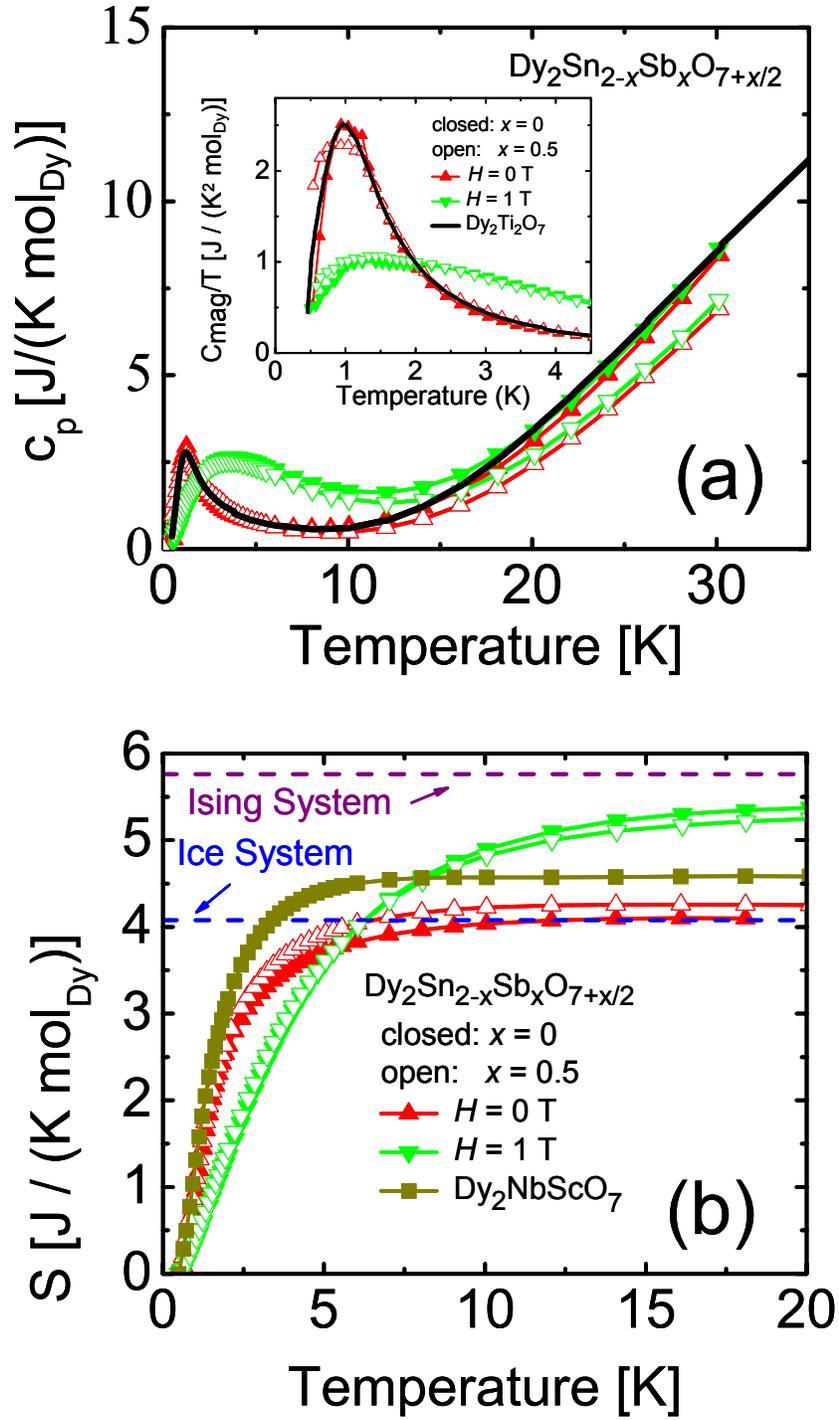